\documentstyle[11pt,aasms4,epsf]{article}
\lefthead{P. Andreani \ea }
\righthead{Ho RXJ0658}

\def\ea{{\it et al.}\ }

\def\comp{Comptonization}

\def\2pr{^{\prime\prime}}

\def\araa{{\it Ann. Rev. Astr. Ap.}\ }
\def\apj{{\it ApJ}\ }
\def\apjs{{\it ApJS}\ }
\def\aj{{\it AJ}\ }
\def\aaa{{\it A\&A}\ }
\def\aas{{\it A\&AS}\ }
\def\mnras{{\it MNRAS}\ }

\newfont{\cm}{cmss10 scaled 1000}
\newfont{\cms}{cmss12 scaled 1000}
\newfont{\cmss}{cmssi12 scaled 1000}
\newfont{\cmsss}{cmss17 scaled 1000}

\begin{document}

\title{The enhancement and decrement of the Sunyaev-Zeldovich effect\\
towards the ROSAT Cluster RXJ0658-5557
\footnote {Based on observations collected with the ESO-Swedish
SEST 15m telescope (La Silla, Chile).}}

\author{P. Andreani\altaffilmark{2}}
\affil{Dip. di Astronomia, Univ. di Padova, Italy}
\author{H. B\"ohringer\altaffilmark{5}}
\affil{Max-Planck Institut f\" ur extraterrestrische Physik, Garching, Germany}
\author{G. Dall'Oglio\altaffilmark{3}}
\affil{Dip. di Fisica, III Univ. Roma, Italy}
\author{L. Martinis\altaffilmark{4}}
\affil{E.N.E.A. TIB, Frascati, Italy}
\author{P. Shaver\altaffilmark{6}}
\affil{European Southern Observatory, Garching, Germany}
\author{R. Lemke, L.-\AA.~ Nyman\altaffilmark{7}}
\affil{European Southern Observatory, La Silla, Chile}
\author{R. Booth\altaffilmark{8}}
\affil{ONSALA Space Observatory, Onsala, Sweden}
\author{L. Pizzo\altaffilmark{3}}
\affil{Dip. di Fisica, III Univ. Roma, Italy}
\author{N. Whyborn\altaffilmark{9}}
\affil{S.R.O.N. Groningen, The Netherlands}
\author{Y. Tanaka\altaffilmark{10}}
\affil{Astronomical Institute, University of Amsterdam, the
Netherlands}
\author{H. Liang\altaffilmark{11}}
\affil{Physics Lab., University of Bristol, United Kingdom}
\altaffiltext{2}{Dipartimento di Astronomia, Vicolo Osservatorio 5, I-35122, 
Padova, Italy. e-mail: andreani@astrpd.pd.astro.it}
\altaffiltext{3}{Dip. di Fisica, III Univ. Roma, P.le Aldo Moro
2, I-00185 Roma, Italy. e-mail: dalloglio@roma1.infn.it,
pizzo@roma1.infn.it}
\altaffiltext{4}{E.N.E.A. TIB, Frascati, Italy}
\altaffiltext{5}{Max-Planck Institut f\" ur extraterrestrische Physik,
Garching, Germany.}
\altaffiltext{6}{European Southern Observatory, Garching, Germany.
e-mail : pshaver@eso.org}
\altaffiltext{7}{European Southern Observatory, La Silla, Chile.
e-mail : rlemke@eso.org, lnyman@eso.org}
\altaffiltext{8}{ONSALA Space Observatory, Onsala, Sweden.}
\altaffiltext{9}{S.R.O.N. Groningen, P.O. Box 800, Groningen, The Netherlands.
e-mail: nick@sron.rug.nl}
\altaffiltext{10}{Astronomical Institute, University of Amsterdam, 1098 SJ
Amsterdam, the Netherlands}
\altaffiltext{10}{Physics Laboratory, University of Bristol, Tyndall Av.,
Bristol BS8 1TL, UK. e-mail: h.liang@bristol.ac.uk}

\begin{abstract} 

We report simultaneous observations at $1.2$ and $2\,${\it mm}, with a
double channel photometer on the SEST Telescope, of the X-ray
cluster RXJ0658-5557 in search for the
Sunyaev-Zel'dovich (S-Z). 
\hfill\break
The S-Z data were analyzed using the relativistically correct 
expression for the \comp and we find
from the detected decrement $(2.60 \pm 0.79 )
~ 10^{-4} $, which is consistent with that computed using the X-ray (ROSAT
and ASCA) observations.
The uncertainty includes contributions due to statistical uncertainty 
in the detection and systematics and calibration.
The 1.2 {\it mm} channel data alone gives rise to a
larger \comp ~parameter and this result is discussed in terms
of contamination from foreground sources and/or dust in the cluster or from a
possible systematic effect.
We then make use of the
combined analysis of the ROSAT and ASCA X-ray satellite observations
to determine an isothermal model for the S-Z surface brightness.
Since the cluster is asymmetrical and probably in a merging process,
models are only approximate. The associated uncertainty can, however,
be estimated by exploring a set of alternative models. We then find
as the global uncertainty on the \comp ~parameter a factor of 1.3.
Combining the S-Z and the X-ray measurements, we determine
a value for the Hubble constant. The 2 {\it mm} data are consistent with
$H_0(q_0=\frac{1}{2})= 53  ^{+ 38} _{- 28} \,{\rm kms}^{-1}{\rm Mpc}^{-1}$,
where the uncertainty is dominated by the uncertainty in
models of the X-ray plasma halo. 
\end{abstract}

\section{INTRODUCTION}

The inverse Compton scattering on the photons of the Cosmic
Microwave Background (CMB) by the hot e$^-$ gas residing in rich clusters
of galaxies causes an effect known as the
Sunyaev-Zeldovich effect (Zeldovich \& Sunyaev 1969;
Sunyaev \& Zeldovich 1972; Sunyaev \& Zeldovich 1981).
The interaction gives rise to a net intensity change with respect to
the undistorted blackbody spectrum which has a distinctive behaviour
as a function of wavelengths: it is negative for wavelengths larger than
$\lambda_0=1.4$ {\rm mm} (decrement) and positive at shorter wavelengths
(enhancement).

The original computation by Sunyaev and Zeldovich of the
net transfer of energy from the hot e$^-$ to the microwave
photons predicts a signal for the
relative temperature change:

\begin{equation}
\left({\Delta T \over T}\right)_{therm}
 = ~y~(x ~{{ e^x +1 }\over {e^x -1}} -4)
\end{equation}

\noindent
where T is the CMB temperature, $x=h\nu/kT$ and $y=\int (kT_e/mc^2)~
n_e \sigma _T d\ell \simeq (kT_e / mc^2) \tau$
is the comptonization parameter, $n_e$, T$_e$, $m_e$ are
the electron density, temperature and mass and
$\sigma _T = 6.65 ~10^{-25} ~~ \rm cm ^{-2} $ is the Thomson
cross section. Equation 1 is an approximate solution of the full
kinetic equation for the change of the photon distribution due to scattering.
A more accurate solution gives rise to corrections, $\Psi (x, T_e) $,
which are not negligible at high frequencies (see Wright 1979; Rephaeli
1995a and 1995b, Challinor \& Lasenby, 1998):

\begin{equation}
\left( {\Delta T \over T}\right) _{therm} = \left(\frac{e^x - 1}{xe^x} \right)
\int d\tau \Psi (x, T_e) 
\end{equation}

If the cluster has a
peculiar velocity relative to the frame where the CMB is isotropic
an additional effect should be measured, which is generally called {\it
kinematic}.
The motion of the gas cloud will induce a Doppler change whose
relative amplitude, $({\Delta T \over T} )_{kin}$, does not depend on the
frequency but only on the peculiar velocity and cloud optical depth
for Thomson scattering, $\tau$: $({\Delta T \over T})_{kin} = -
{v_r \over c} \tau $
(where the minus sign refers to a cluster receding from the observer).
Since both effects are very small the net relative temperature change
is just the sum of the two.

The combination of X-ray observations of the thermal bremsstrahlung
emission by the hot gas and the radio and millimetric data of the clusters
is a powerful cosmological tool to investigate physical processes
in the earlier universe, the determination of H$_0$ and q$_0$, the peculiar
velocities of clusters and the nature of the intracluster medium (ICM) (see the
original papers by Zeldovich \& Sunyaev).

Many observational efforts, mostly in the Rayleigh - Jeans (R-J) part
of the spectrum, were carried out
to detect these effects (see e.g., the recent reviews by Rephaeli 1995a
and by Birkinshaw 1997). The expected decrement at
centimeter wavelengths is found
towards A2218, A665, 0016+16, A773, A401, A478, A2142, A2256 and Coma
(Birkinshaw \& Gull 1984; Birkinshaw M. 1991;
Klein et al. 1991; Jones et al. 1993; Grainge et al. 1993;
Herbig et al. 1995; \cite{C96}; Myers et al. 1997)
and at 2.2 {\rm mm} towards A2163 (Wilbanks et al 1994).

Measurements near the Planckian peak and on the Wien side are in
principle  more
attractive since: (a) they allow the spectrum to be uniquely identified
as SZ (as opposed to primordial CMB fluctuations),
(b) the intensity enhancement relative
to the Planckian value is larger than the magnitude of the R-J decrement;
(c) sources in the cluster are expected to give a negligible contribution at
high frequency; (d) they allow a measurement of cluster peculiar velocities.
Clearly an unambiguous signature of its presence is provided by
the simultaneous detections of the enhancement and the decrement.
\hfill\break
Recently Holzapfel et al. (1997)
towards A2163 have reported
the first detection at millimetric wavelengths of both the decrement and
the enhancement of the effect, while Lamarre et al. (1998) reported detections
of the same cluster at 630 $\mu m$ and 390 $\mu m$ of the increment.

In this paper we report a combined analysis {\rm mm}, ROSAT and ASCA
observations towards the ROSAT Cluster RXJ0658-5557, for
which the simultaneous observations of the effect
at 1.2 {\rm mm} (positive) and at 2 {\rm mm} (negative) were
preliminary reported
elsewhere (Andreani et al. 1996a). The cluster was chosen because of
its hot temperature ($T_e \sim $ 17 keV), high X-ray luminosity
($L_x = (3.5 \pm 0.4)  \cdot 10^{45} h_{50}^{-2}$ erg s$^{-1}$
in the range 0.1 - 2.4 keV) and large distance ($z=0.31$),
this last property is needed to cope with the SEST beam size of
44$^{\prime\prime}$ on the sky (see below).
\hfill\break
Sections 2 and 3 describe {\rm mm},  ROSAT and ASCA observations,
while the implications of
these measurements for the evaluation of the Hubble constant
are discussed in \S 4. In section 5 we discuss the contamination from
foreground and/or intracluster sources.

\section{MILLIMETRIC OBSERVATIONS}

\subsection{ The Instrument}

A double channel photometer was built and devoted to the simultaneous
search for the
enhancement and decrement of the S-Z effect. The system works simultaneously
at 1.2 and 2 {\rm mm} using two bolometers cooled at 0.3 K
by means of a $^3$He refrigerator. The 2 {\rm mm} band includes the peak
brightness of the decrement in the
S-Z thermal effect, while the 1.2 {\rm mm} bandwidth is a compromise between
the maximum value of the enhancement in the S-Z and the atmospheric
transmission. 
The wavelength ranges are defined by two
interference filters centred at 1.2 and 2 {\rm mm}
cooled at 4.2K with bandwidths 350 and 560 $\mu m$
respectively (response curves of the optical trains can be found
in \cite{PIZ} and in the appendix A). The collecting optics are made up of
a dichroic mirror splitting the incoming radiation into two f/4.3 Winston
Cones cooled at 0.3 K, which define a field of view on the sky
of 44$^{\prime \prime}$ at both frequencies. The beam separation
on the sky was limited by the antenna chopping system and was set to the maximum
chopping amplitude, 135$^{\prime \prime}$.
The pointing accuracy was frequently checked and was always better then 3-4$^{\prime \prime}
$. Alignment of the two beams was accurately determined with calibrators
and turned out to be better than 2$^{\prime \prime}$.
We tested the quality of the beam shapes and measured the beam widths
imaging the planets. Beam widths are reported in table 1. Any
deviation from a gaussian surface
turned out to be less than 1 \%  (see \cite{PIZ} for details).

This photometer was adapted to the focus of the SEST
and its performance was tested during an observing run
in September 1994. Details of the instrument can be found
in Pizzo et al. (1995) and \cite{A96b}, a brief description is also reported in
appendix A.

\subsection{The Observations}

Responsivity measurements were conducted regularly during September 1994 and
1995 by observing planets (Uranus, Saturn, Jupiter, Venus and Mars) with
Jupiter used as primary calibrator. The main figures measured at the focus
are listed in table 1. The reported NET is given in antenna temperature, while
sensitivities are expressed as relative change of the
thermodynamic temperatures in one second integration time
(see \cite{PIZ}).

Calibration uncertainties are mainly related to those of the planet
temperatures at {\it millimetric} wavelengths. We took the values quoted by
Ulich (1981 and 1984) who took measurements of the planets at the same
frequencies as we did. The uncertainties on the planet brightness quoted by
this author are less than 10 \%. However, different measurements taken with our
 instrument during different nights of observations
produce sometimes results
off by more than 10 \%, due to changes in the observing conditions.
Thus we conservatively estimate
that the final uncertainties on the antenna temperature is of $\sim$ 15 \%.

A total integration time of 15000s was spent on the source in several
different nights during September 1995
and the same integration time was spent on blank sky
located 15m ahead in right ascension. Some of the data
were discarded for the
present work because they were corrupted by poor weather conditions. The
remaining data were those collected during the nights when the sky opacity was
very low ($\tau_{1mm} < 0.1$ with an average value of
$<\tau_{1mm}> = 0.07$, $\tau _{2mm} < 0.05$
with an average value of $<\tau _{2mm}> = 0.03$) and the sky emission
very stable thus producing a very low sky-noise.
The {\it effective} integration time was 12000s on-source.

In order to get rid of the major sources of noise in this kind of experiment,
fluctuations in the atmospheric emission and systematics from the
antenna, the observing strategy makes use of two combined procedures:
the common three-beam technique, beam-switching + nodding, which gets
rid of the linear spatial and temporal variations in the atmospheric
emission, and the observations of blank sky regions located 15m ahead in right
ascension ($\Delta \alpha = 15 m$)
with respect to
the source. This latter implies that for each 10m integration on-source
(10m integration + overheads give a total tracking time of 15m) a similar
integration is performed on blank sky.
This level of switching corrects for offsets dependent on telescope
orientation and related to diffracted radiation from the environment.
The comparison between the two signals provides
a measurement of the systematics introduced by the antenna. In fact,
the instrument tracks the same sky position twice with respect to the local environment
once on-source and the other on the blank sky.
The choice of 600s of integration on and off the source is a compromise
between the minimization of the time wasted on overheads and the need of
minimizing the atmospheric variations between one observation and the other.
In this way the efficiency of the observations is not greatly reduced by
frequent slewing, but small enough that temporal variations are not too
severe.\hfill\break
Note that the sequence on-source and off-source (on blank sky) was
performed by the computer controlling the antenna and was therefore entirely
automatic. The synchronization between the two measurements was
as high as possible, with an estimated error of less than 2 s, negligible with
respect to the beam size and integration step (10 min).

As the telescope tracks the cluster across the sky the reference beams trace
circular arcs around the cluster. The reference beams are always separated
from the on-source position only in azimuth and their positions 
are given in terms of parallactic angle $p$ (the angle between the
North Celestial Pole and the zenith):

$$ \tan p = { \cos \phi \sin H \over {\sin \phi \cos \delta
-\cos \phi \sin \delta \cos H}} $$

\noindent
where $\phi$ is the geographic latitude, 
$H$ and $\delta$ are the hour angle and the declination of the object.
It would be better to have a wide distribution
of parallactic angles to avoid contamination from sources in the reference beams.
We overplot in figure 6 the positions of the main and the circular
arcs swept by the
reference beams on the S-Z map obtained by the X-ray image
(see below for details).
Because of the complex X-ray map and the
limited beam throw which results in reference beams being not completely
out the X-ray emission, we need to carefully model the expected signal with
this configuration before comparing theoretical expectations with the
real data. This point is further discussed in \S 4.

The data analysis procedure is described in detail in \cite{A94} and
\cite{A96b} 
and here only the basic methods are described.
The data consist of 200s integration blocks (hereafter called {\it one
scan}) each containing 20 {\it 10s subscans} taken at two different
antenna positions, A and B, with the reference beam on the left and
right respectively. The integration sequence
was ABBAABBAABB...A. A 2$^{nd}$ order
polynomial fit was subtracted from each scan in order to get rid of
offsets in the electronics and atmospheric large scale trends not
completely cancelled out by the three-beam switching. From each subscan
spikes due to equipment malfunction were removed (but the
fraction of rejected data is less than 1\%) and
very high-frequency atmospheric variations were smoothed with the 
Savitzky-Golay filter algorithm (Press \& Teukolsky 1990).
For each
subscan a mean value for the differential antenna temperature,
$T_A$ or $T_B$, is found by averaging over the 10s.
The variance, $\sigma^2_{A,B}$,
is estimated with a procedure of bootstrap re-sampling (Barrow et al 1984).
This method is widely used to take into account correlations among data
(induced in this case by the filtering).
The procedure requires for each subscan the creation of many mock
{\it subscans} (in this case 1000 each), each containing the same number of
data as the real one, by randomly redistributing the data, i.e. the sequence 
number of the data is randomized. This means that in some cases one datum 
can be considered more than once, while in some other cases it can be 
discarded. A mean value for the differential antenna temperature 
is then computed for each mock subscan. The {\it bootstrap}
uncertainty for each real sub-scan is estimated by the standard deviation of
the mean value of the 1000 mock subscans.

The signal is then obtained by subtracting each couple of subscans
 $\Delta T_i = {{T_A - T_B }
\over {2}}$ with variance $\sigma ^2 _i = {{ \sigma^2_A + \sigma^2 _B}
\over {2}}$ 


\noindent
Weighted means are computed for each 200s integration (1 scan)
on each sky position (when the antenna tracked the source,
hereafter called $\Delta T_{ON}$, and when the antenna tracked the
blank sky, hereafter called $ \Delta T_{OFF} $). Cluster signals 
are then estimated from the subtraction of each off-source from each
on-source:
$ \Delta T_{SZ} = \Delta T_{ON} - \Delta T_{OFF} $
and the quadratic sum of the two standard deviations are
used to estimate errors:
$ \sigma _{SZ} ^2 = \sigma ^2 _{ON} + \sigma ^2 _{OFF} $.

Figure 1({\it a,b}) shows in the lower panels these differences
in {\it antenna temperature} as a function of time for both channels.
The solid lines represent the maximum likelihood estimates of the 
$ \Delta T_{SZ} $, while dotted lines correspond to $\Delta T_{SZ} \pm 3\sigma$.

The $\chi ^2$-tests, performed over the data reported in Figure 1,
give the following values $\chi ^2=37.2$ and $\chi ^2=34.2$
for 50 degrees of freedom at 1 and 2  {\rm mm} respectively.
From the maximum likelihood estimates we find $\Delta T_{1mm} = + 0.30 \pm 0.07 $
 mK, $\Delta T_{2mm} = -0.50 \pm 0.15 $ mK,
where the uncertainties are given by the $70\%$ confidence range.
This was found by estimating the width of the likelihood curve
corresponding to the values of the signal where the likelihood drops
by a factor of $1.71$ from its maximum. Maximum-likelihood curves are
shown in figure 2.

Converting these values to thermodynamic
ones we find: 0.96$\pm$0.22 mK at 1.2 {\rm mm} and 
-0.86$\pm$0.26 mK at 2 {\rm mm}. The errorbars associated with these values
are only due to the {\it statistics}. To estimate those due to
observing systematics we have proceeded as follows. 

To test any position-dependent systematics,
Figure 1 ({\it a,b}) shows also in the upper panel the sum
$ \Delta T_{sys} = \frac{\Delta T_{ON} + \Delta T_{OFF}}{2} $. The statistics in
these cases provide as weighted averages: 0.12$\pm$0.08 mK
at 1{\rm mm} and 0.04$\pm$0.13 mK at 2 {\rm mm}. The associated errorbars
will be quoted in the following as systematics uncertainties and will be
considered 
in the final values of the comptonization parameter (see \S 4.3).

Another way to take into account of the systematics is to estimate
the errorbars as in \cite{A94} and \cite{A96b}.
For each $\Delta T_{SZ}$ a weight is assigned: $ w_m = (\sigma^2 _{SZ} +
\sigma^2 _P)^{-1}$, where $\sigma ^2 _{SZ}$ represents the variance due
to short-term fluctuations (those reported in figure 1 associated with each
 on-source/off-source difference), while $\sigma ^2 _P$ that due to medium-term
variations, i.e. that related to any long-term systematics in these
differences.

The final {\it weighted} mean is given by:

\begin{equation}
\langle \Delta T _{SZ} \rangle =
{{ \sum _{m=1} ^N \Delta T_{SZ} \cdot w_m } \over
{ \sum _{m=1} ^N w_m }}
\end{equation}

\par\noindent
with estimated variance:

\begin{equation}
\sigma ^2 _f = \frac{1}{N-1}
{{ \sum _{m=1} ^N (\Delta T_{SZ} - \langle \Delta T_{SZ} \rangle) ^2 \cdot w_m }
\over { \sum _{m=1} ^N w_m }}
\end{equation}

\noindent
In this case one finds: $0.35\pm0.11$ and $-0.61\pm0.20$ mK ({\it antenna temperature})
at 1 and 2 {\rm mm} respectively. The errorbars here equal the quadratic sum
of that due to statistics and that due to systematics, estimated above.

In order to check whether the detections are spurious due to the
equipment (microphonics and slewing of the telescope),
we have carried out additional measurements
blocking off the beam: an aluminum sheet was put on
the entrance window of the photometer, located in order to cover the entire
window and to prevent that diffracted radiation entering the photometer.
The same double blank-sky observing sequence 
on and off the source, i.e. the same
alto-azimuthal paths, were tracked with this configuration for 5000 s
and the
resulting signals are shown in figure 3. 

No signals are detected with the window covered by the metal sheet and
there is no trend in the signal as a function of the alto-azimuthal position.
We conclude that there are no systematics coming from the photometer, the
two channels behave
properly and do not introduce spurious signals.
The r.m.s. values of the signals stored with the window covered by the metal
sheet turn out to be 0.07 and 0.15 mK (antenna temperatures)
at 1 and 2 {\rm mm} respectively. These values can be used as
an estimate of the instrumental noise during the observations once
scaled to the integration time
spent on the source: 0.05 and 0.1 mK (antenna temperatures)
at 1 and 2 {\rm mm}, respectively.\hfill\break
This does not mean that this procedure takes into account all the
antenna systematics, when the  photometer looked at the source and at
the blank sky. Some spurious signals may survive
if the antenna did not  precisely track the same paths relative to the
local environment, because of a loss of synchronization.
As already reported above we have tried to
quantify the error in the position
of the hour angle track for the off position, i.e. the error in
synchronization and we estimate that it is of 2s, which is
much lower than the beam size and the nodding interval of 10s.

A further check on environmental systematics, which could affect the S-Z
signals,
was performed by investigating any dependence of blank-sky signals on elevation.
Figure 4 shows the signals recorded when the
antenna pointed the blank sky as a function of the elevation.
There is no evident systematics affecting the observations.

\section{ROSAT AND ASCA OBSERVATIONS}

Details of ROSAT and ASCA observations of this source can be found in
B\"ohringer and Tanaka (1997). Here we briefly summarize the main
X-ray properties.

\subsection{ROSAT data}

The galaxy cluster RXJ0658-5557 was independently discovered in
the EINSTEIN slew survey (\cite{TUC}) as the source 1E0657-55
and in a follow-up identification program for galaxy clusters in the 
ROSAT All Sky Survey (RASS) at ESO started in 1992 
(\cite{BO94}, \cite{GUZ}) as X-ray source RXJ0658.5-5557.
The cluster is extremely X-ray bright and was therefore
 scheduled for various follow-up observations with ROSAT, ASCA and optical
spectroscopy measurements.

RXJ0658-5557 was detected in the RASS with a count-rate of $0.5 \pm
0.04$ cts s$^{-1}$ in the 0.1 to 2.4 keV band and $0.41 \pm 0.04$
cts s$^{-1}$ in 0.5$\div$2 keV  which corresponds to a 0.1 - 2.4 keV flux of
$(8.5 \pm 0.9) \cdot 10^{-12}$ erg $s^{-1}$ cm$^{-2}$ and
an X-ray luminosity of $L_x = (3.50 \pm 0.40)  \cdot
10^{45} h_{50}^{-2}$ erg s$^{-1}$ (0.1 - 2.4 keV).
For the luminosity
calculation we used a value of $17$ keV for the temperature,
$N_H = 3.5 \cdot 10^{20}$ cm$^{-2}$ for the absorbing column density
and a metallicity of 0.35 of the solar value, parameters inferred from the
ASCA spectral analysis (see below).
The cluster was also observed in a pointed observation with the
ROSAT HRI (PI: W.H. Tucker) for 58 220 s and was detected
with a count rate of $(0.166 \pm 0.005)$ cts s$^{-1}$  yielding and 
X-ray luminosity of $L_x = (3.44 \pm 0.15)  \cdot 10^{45}  h_{50}^{-2} $ 
erg s$^{-1}$ in the rest frame 0.1-2.4 keV energy band in very good
agreement with the PSPC data.
The ROSAT band flux calculated from the 
HRI observation is $(8.3 \pm 0.3) \cdot 10^{-12}$ erg $s^{-1}$ cm$^{-2}$.  
We also made use of a ROSAT PSPC observation conducted in
February 1997 with an exposure of about 5 ksec to check the astrometry. 

\noindent
Figure 5 shows the ROSAT HRI image of the cluster: the X-ray
morphology is quite complex, the cluster center being
dominated by two blobs with the western component (hereafter component 2)
less luminous. It
may indicate a process of merging of two substructures, a morphology
which complicates the modeling of the X-ray surface brightness as
discussed in \S 4.

\subsection {ASCA Observations}

RXJ0658-5557 was observed with ASCA on 10 May, 1996, for a total
observing time of $\sim 40$ ksec. 
A peak at around 5 keV is conspicuous due to the
redshifted iron K-lines from which the redshift parameter can be
determined. The Raymond-Smith model was employed for fitting model
spectra to the observed data. The best-fit results obtained from
the GIS spectrum are $kT = 17.0 \pm 4.0$ keV, the abundance of
elements $0.35 \pm 0.13$ of the solar abundance, the redshift,
$z = 0.31 \pm 0.03$, and the absorption column $N_H
\le 6 \cdot 10^{20}$ cm$^{-2}$. The errors are 90\% confidence.
The redshift of the cluster was also confirmed via optical spectroscopy
of eight galaxy members (data are published in Tucker et al. 1998). 

The observed flux in the 2 - 10 keV range is approximately $0.9 \cdot 
10^{-12}$ erg s$^{-1}$ cm$^{-2}$ from which the total bolometric 
flux is estimated to be $\sim 2 \cdot 10^{-11}$ erg s$^{-1}$ cm$^{-2}$.
For z=0.31, the bolometric luminosity is $\sim 1.1 \cdot 10^{46}$ erg 
s$^{-1}$. Thus, RXJ0658.5-5557 is one of the hottest and most luminous 
clusters known to date, hence ideal for observations of the Sunyaev
and Zeldovich effect.

The accuracy of the absolute calibration of the two Satellites is difficult to
assess. Cross-calibrations among the different instruments agree within a
factor of 10 \%. 

\section{MODELING THE HUBBLE CONSTANT}

The angular diameter distance to a cluster can be observationally determined
by combining measurements of the thermal S-Z effect and X-ray measurements
of thermal emission from intracluster (IC) gas. Thus the value of H$_0$
can be deduced
from these measurements, as many authors have discussed in the past
(\cite{Cal79}; \cite{BHA};
Holzapfel et al. 1997; \cite{M97} ).

\subsection{Modeling the X-ray data}

\noindent
Unfortunately this cluster has a complex morphology,  making
it difficult to use a simple geometrical model to fit the
surface brightness.
There is not a unique way to de-project the X-ray surface
brightness distribution to derive the volume emissivity and the
electron density distribution. A pragmatic approach is to model the
cluster with two symmetrical components centered at the two X-ray
maxima. While the smaller Western component (2) is quite round and compact,
the Eastern one (component 1) appears elongated in the North-South
direction in the central region. At large radii, however,
the cluster becomes more
azimuthally symmetrical. Thus a spherically symmetrical model provides a fair
average approximation also to the slightly elongated structure as
shown by e.g. Neumann \& B\" ohringer (1997). We then fit
empirical beta models to the azimuthally averaged surface brightness profiles
and compute through an analytical de-projection 
the distribution of the IC gas column density along the line of sight
(\cite{Cal76}):

\begin{equation}
n(\theta) \propto \left(1+\frac{\theta^2}{\theta_c^2}\right)^{- \frac{3}{2}
\beta + \frac{1}{2}} ,
\end{equation}

\noindent
in which $\beta$ and the cluster angular core radius ($\theta_c$) 
are left as free parameters.
We performed the actual fits to the profiles of the undistorted side
of the component 1 and the less distorted western side of component 2 in the
high resolution ROSAT/HRI data. The resulting parameters of the two best-fit
models are listed in table 2.

The electron densities corresponding to these surface brightness values
are $ 6.3 ~ 10^{-3}$ and $1.5 ~10^{-2} ~~ cm^{-3}$ for the Western (2) and
Eastern (1) component, respectively ($H_0=50 km s^{-1} Mpc ^{-1}$ 
was used in this calculation).

The X-ray observations were then used to make a prediction for the
expected S-Z increment or decrement. The numerical
values in this calculation depend on the Hubble parameter, $H_0$, and
in this way by comparing the predicted with measured
values an estimate of the absolute distance of the cluster and $H_0$
can be made.

The first step was to compute the Comptonization parameter, $Y$, for
the models given in the previous subsection. The parameter $Y$ at the
cluster centre is given by:

\begin{equation}
Y = \frac{k T_e \sigma _\tau}{m_e c^2} \int^{+R} _{-R} n_e (r) dr
\quad h_{50} ^{-1/2}
\end{equation}
 
\noindent
where $R$ is the outer radius of the cluster. We took the estimated
virial radius, $3 h_{50}^{-1} Mpc$.

Since we do not have spatially resolved information on the
temperature distribution, the thermal structure of
the IC gas is also assumed to be radially 
symmetrical and isothermal. One effect of deviations
from isothermality is discussed below.


Numerically a $\beta$-modelled cluster has a \comp ~parameter:

\begin{equation}
Y = 1.3 ~10^{-27} ~cm^2 \left(\frac{T_e}{1~keV}\right) ~n_{e}(0) r_c
\left(\frac{\sqrt{\pi} \Gamma(1.5\beta -
0.5)}{\Gamma(1.5\beta)}\right) h_{50}^{-1/2}
\end{equation}

\noindent
where $n_e (0)$ is the central electron density and $r_c$ is the core
radius of the cluster plasma halo.
The last equation is not {\it formally} consistent with eq. (4)
because in this case the integration limit is infinite
while in the former equation the integration is performed up
to the virial radius. The error on the $Y$ parameter
introduced by choosing a wrong (because
unknown) outer radius is estimated to be small (less than 5 \%).

\noindent
Inserting the values listed in table 2 we find the following
$Y$-parameters at the center of each component:

\begin{equation}
Y = 4.90 ~10^{-4} h^{-1/2} _{50} \quad {\rm for ~component ~1}
\end{equation}
\begin{equation}
Y = 4.80 ~10^{-4} h^{-1/2} _{50} \quad {\rm for ~component ~2}
\end{equation}

To compare the predicted S-Z effect values with those obtained from the
observations, the $Y$-parameter at the position of the observations in
the composite model must be evaluated.
Since
the observed S-Z signal is the true one modified by the telescope
primary beam and the beam switching, the predicted $Y$-parameter must be
convolved with beam pattern and throw. We can well approximate the beam
pattern with a Gaussian (\cite{PIZ}):

\begin{equation}
G(\alpha) = {1 \over 2\pi \sigma ^2} exp ( - {\alpha ^2 \over 2 \sigma ^2})
\end{equation}

\noindent
$\sigma $ is roughly half of the beam width ($44^{\prime \prime}/2.3$ ) 
and the average signal in the Gaussian beam is:

\begin{equation}
\tilde Y(\rho) = 2 \pi \int \rho^\prime  d\rho^\prime
 {1\over 2\pi L^2} exp (-{\rho^\prime}^2/2L^2) Y(\rho^\prime) 
\end{equation}

\noindent
where $L=D_a \sigma$ and $\rho$ is the angular radius from the cluster center.
Once $\rho$ is expressed in terms of R.A. and declination the integral
becomes a double integral.

\noindent
The beam switching introduces a gradient in this value:

\begin{equation}
 < \Delta Y_{sw} > = \tilde Y (0) - {1\over 2} \tilde Y(R_{-})
- {1\over 2} \tilde Y (R_{+})
\end{equation}

\noindent
where $R_{\pm} $ corresponds to the two positions of the reference beams
at a distance from the center, $D=D_a \theta _D$, where $\theta _D
= 135^{\prime \prime}$ is the beam separation.

\noindent
To compare the measured values with the expected signals from the cluster
we have first convolved the predicted S-Z surface brightness from the
X-ray model using a two-dimensional filtering (equation 8) and then
averaged the $Y$ values over the positions of the projected reference
beams in the north-western and south-eastern arcs,
whose average centers are given by the coordinates (J2000): 104.6665d,
-55.9231d (northern offset), 104.6063d, -55.9865d (southern offset).
The central beam position is 104.6363d, -55.9551d.
The resulting values are:

\begin{equation}
6.4 ~10^{-4} ~(C) ~~ 4.0 ~10^{-4} ~(N) ~~ 3.3 ~10^{-4} ~(S)
\end{equation}
 
\noindent
where $C$, $N$ and $S$ stay for {\it center}, {\it north} and {\it
south} respectively. The averaged $\Delta Y$-parameter is: $\Delta
Y = 2.70 ~10^{-4} $.

At this point it is very important to investigate the influence
of the pointing accuracy on the results. On the one hand, there is
the uncertainty on the pointing offset of the telescope (discussed 
in \S 2); on the other hand, there are those related to
the PSPC and HRI pointings.
While we can safely assume that the {\it relative} pointing accuracy of
SEST is better than 3$^{\prime \prime}$, the recovering of the
astrometry of the PSPC image was not an easy task, because of the
uncertainties in the positions of the known sources in the field.
\hfill\break
The PSPC and HRI pointings agree within 10$^{\prime \prime}$.  
Therefore we allow an overall pointing accuracy of about 20$^{\prime \prime}$. 
This reflects into an uncertainty on the estimated $\Delta Y$.

Figure 6 shows the convolved S-Z surface brightness predicted with the
X-ray model described above. Superimposed are the northern
and southern arcs of the reference beams (crosses) and the
position of the main beam (open square).

We see that the combination of the beam size and the limited chop
throw has the effect that the \comp ~parameter has about half of
the central value at the offset reference positions.

\subsection{Errors on $Y$ from X-ray modeling}

A critical point in these results is the uncertainty on
the idealized
model of the cluster plasma halo. In particular,
the merging configuration, indicated by the high resolution X-ray 
image of the cluster, does not allow a correct modeling.  
There is, however, a clear indication in the X-ray images 
that the main component is quite extended and almost symmetric 
at large radii while the smaller component appears bright 
due to its high central density but is quite compact. We 
can therefore conclude that the main component carries 
a very large fraction of the gas mass and total mass 
and is only slightly disturbed on a global scale by the merging
smaller and more compact group. We expect that 
the major contribution to the SZ signal comes by far 
from the main component and that the small infalling  
cluster makes only a minor contribution. The strategy to 
measure the SZ signal slightly off from the peak at  
the opposite side of the merger further decreases the influence 
of the infalling clump. 
               
Thus a pragmatic approach to estimate the uncertainty of 
the result for the Comptonization parameter is to explore several alternatives,
preferably extreme models, which should help to bracket the true
result. 

\begin{itemize}

\item
The first alternative model considers only the halo
of component (1) to calculate the expected value for
the Comptonization parameter. It turns out that $\Delta Y
= 2.2 ~10^{-4}$. The justification for this model is 
that the smaller component may actually have 
a much smaller $T_e$ than the main component and/or it may not
be as extended as assumed in our model. In both cases the
effect of this component would be considerably overestimated.
Thus by completely neglecting this latter component the
resulting $Y$-value is a lower limit to the real one.


\item
We then compute another alternative model changing the temperature profile
from the isothermal case. If there is a temperature gradient decreasing
with radius, the $Y$-image will be more compact, at variance with what we found
using the second model. The 
easiest analytical model for a temperature gradient is a polytropic
model with
\begin{equation}
\frac{T(r)}{T_{0}} = \left(\frac{n(r)}{n_{0}}\right)^{\gamma -1}
\end{equation}

where $\gamma$ is the polytropic index. If we assume a quite large 
value $\gamma =1.4$, we find a value for $\Delta Y = 2.10 ~10^{-4}$.
The resulting difference in $Y$ is, on the one hand, enhanced from the
greater compactness of the cluster and, on the other hand, shrunk
because of the decrease of the central value, $Y_{0}$.

\end{itemize}

Comparing the results for the various models we quote as a conservative
uncertainty a factor of 1.3 in the parameter $\Delta Y $. 
Since $H_0$ depends quadratically on
this value (equation 4), the uncertainty 
in the final estimate of $H_{0}$ is roughly a factor 1.6.
\hfill\break
The uncertainty in the overall temperature measurement is
less than 4 keV at the 2$\sigma$ limit and introduces an error
of 25\%.

Another aspect of the structure of the intracluster plasma 
that introduces uncertainties is the possible clumpiness of this medium. 
This effect was, for example, considered by Birkinshaw (1991)
and Holzapfel et al. (1997)
 and investigated by means of simulations by  \cite{ISS} and \cite{RSM}.
For reasonable clumping scenarios, a clumpy medium has as effect
a lowering the H$_0$ value by about 10 - 20\%.
No direct evidence for a clumpy intracluster medium 
was found, however, and the models suggested are still 
highly speculative as well as the following estimated uncertainty.

We then consider 65\% as a conservative uncertainty on the estimate of the
Hubble constant quoted below derived only from modeling.

\subsection{Modeling H$_0$}

The conversion of the \comp~ parameter to the observed change in
antenna temperature is given by equations 1 and 2, whose right-hand side
must be integrated over the instrumental band-widths. Following \cite{CL}
we integrate over the instrument bandwidths equation 2,
which takes into account the relativistic corrections, and find:

\begin{equation}
\left(\frac{\Delta T}{T}\right) _{1.2mm} = 0.38 ~Y
\end{equation}

\begin{equation}
\left(\frac{\Delta T}{T}\right) _{2mm} = -1.20  ~Y
\end{equation}

\noindent
Hence the predicted versus observed changes are:
\begin{equation}
\left(\frac{\Delta T}{T}\right) _{X-ray} = 1.03 ~10^{-4} h ^{-1/2} _{50}
\quad \quad
\left(\frac{\Delta T}{T}\right) _{obs} = 3.3 ~10^{-4}
\end{equation}

\noindent
for the 1.2 {\rm mm} channel, giving a formal result of the Hubble
constant of $H_0=5 ~km~s^{-1} ~Mpc^{-1}$.
For the 2 {\rm mm} channel, one finds:

\begin{equation}
\left(\frac{\Delta T}{T}\right) _{X-ray} = -3.24  ~10^{-4} h ^{-1/2} _{50}
\quad \quad
\left(\frac{\Delta T}{T}\right) _{obs} = - 3.16 ~10^{-4}
\end{equation}

\noindent
giving a formal result of the Hubble
constant of $H_0=53  ^{+ 38} _{- 28}~km~s^{-1} ~Mpc^{-1}$.
The errorbars take into account the uncertainties on the pointing position
of the SEST and ROSAT Telescopes, on the X-ray modeling (as discussed in
\S 4.2) and on the {\rm mm} observations (both statistics and systematics).

\section{Origin of the detected signals}

In the following we discuss the possible contamination to the observed
signals from spurious sources. Radio sources would affect more the 2 {\rm mm}
signal, while thermal sources the 1.2 {\rm mm}.

Let us assume for a while that the decrement seen at 2 {\rm mm} (eq. 16) is
due to the
thermal S-Z effect. From this value the {\it expected} 1 {\rm mm} signal
can be easily computed from eq. 13. Comparing this with the value derived from
the observations (eq.15) a difference of a factor of 3 is found and the
reason of this discrepancy was investigated in several ways, as discussed below.

On the other hand, if we consider the 1 {\rm mm} signal as only due to the
thermal S-Z effect the expected absolute 2 {\rm mm} value
would be higher by a factor of 3. In this case it is straightforward to check
whether any radiosource is present in the main beam.

\subsection{Possible source contamination}

The presence of a source contaminating the signals was checked
as shown below.
We then first assume that a thermal source is present in the main
beam, giving rise to a signal of 0.3 mK in antenna temperature. A
thermal source is stronger at 1.2 {\rm mm} than at 2 {\rm mm} by a
factor of roughly $(2/1.2)^{3.5}$ (\cite{AF}) and therefore
it would be a factor of 6 weaker at 2 {\it mm}:
the expected signal in the 2 {\it mm} channel would be of
only 0.05 mK, well within the errorbars. We then checked the likely
presence of such a source in the cluster center:

(1) No sources in the IRAS Faint Source Catalogue
 are present at the position of the main beam
and/or the reference beams.

(2) If we scale the 60$\mu m$ IRAS flux limit of 240 mJy at 1.2{\it mm} by using
the average flux of nearby spirals (\cite{AF}) we find
that a normal
spiral would give rise to a signal not larger than 0.02 mK in
antenna temperature. So it should be a peculiar galaxy.

(3) If we assume a contribution of many unresolved sources fluctuating in
the beam and take the estimation made by Franceschini et al. (1991) the
expected signal will be not larger than 0.02 mK.

(4) Irregular emission from Galactic cirrus can also give origin to a
signal at these wavelengths. If we take
the estimation by Gautier et al. (1992) and extrapolate the 100$\mu m$ flux
at 1.2 {\it mm} using the average Galactic spectrum a maximum signal of 0.02 mK
is found.

(5) The 60 and 100 $\mu m$ emission from a handful of nearby clusters detected
by IRAS was interpreted as evidence for cold dust in the ICM
(\cite{STIC},\cite{BREG}). Dust grains hardly survive in the ICM 
and their lifetime would be very short in these environments, thus the
probability of its detection  is very low. This does not exclude the
presence of dust in RXJ0658-5557 which would affect 
more strongly the 1.2 {\it mm} channel. 

(6) Let us now consider the case of radio source contamination. In general
radio sources have spectra that decreases with increasing frequency
thus the effect of radio source contamination at 1.2 and 2 {\it mm} are
greatly reduced compared with observations at radio
frequencies. However there are radio sources with spectra that rise
with frequency.  Radio maps of the cluster were obtained with the
Australia Telescope at 8.8 GHz, 5 GHz, 2.2 GHz and 1.3 GHz. The
sources in the 8.8 GHz image were weaker than 10mJy (after primary
beam correction) and all the sources within the primary beam ($\sim
3^{'}$ radius) had spectra that decreases with increasing frequency. A
radio halo source the size of the X-ray emission was found to coincide
with the X-ray emission with a total flux of $\sim 3$mJy at 8.8 GHz
and a spectral index of $-1.7$. The contamination to the main beam is
$\ll 0.001$mK at both 1.2 and 2mm and thus negligible. The reference
beams are not contaminated by the radio sources. Details of the
properties of the radio sources in the field will be the subject of
another paper (Liang et al., in preparation).

We conclude that it is possible that part of the signal at 1.2 {\it mm} is
due to diffuse emission from an eventual
intracluster dust but this
point deserves further investigations and it is the goal of future
researches (for instance ISO mapping at 200 $\mu m$).

\subsection{CMB Anisotropies and Peculiar Velocity}

Peculiar velocities of a cluster could alter the thermal S-Z effect by
a factor $(v_r/c) ~\tau$ (where $\tau$ is the optical depth through
the halo gas). $\tau$ was estimated from the X-ray measurements and turns out
to be $\tau = 39 Y$. 
If the cluster recedes from us with a velocity of 1000
km/s the 1.2 {\it mm} signal would be enhanced only by 10 \% and
will be therefore still incompatible with the 2 {\it
mm} signal. Values of the peculiar velocities  larger than these 
are not found in optical searches and seem to be excluded in most of
cosmological models. 

If part of the signal is due to CMB anisotropies at these scales,
it will be hard to disentangle them from the S-Z kinematic effect
since this latter has a spectrum identical to that of the anisotropies
(see e.g. Haehnelt \& Tegmark, 1995). However, CMB anisotropies originated
at redshifts larger than that of the cluster can be amplified by the
gravitational lensing effect of the cluster itself and affect the signal
quite significantly (see Cen 1998).
\hfill\break
CMB anisotropies will enhance {\it both} signals equally, thus reducing the
amplitude of the 2 {\rm mm} decrement. This means that the real 2 mm signal
due to the S-Z thermal effect alone would be larger 
and the difference between the 1 and 2 mm values would be shrinked. 

\section{Conclusions}

Observations at millimetric wavelengths 
of the Sunyaev-Zel'dovich effect towards the X-ray cluster
RXJ0658-5557  were performed
with the SEST Telescope at La Silla (Chile) equipped with
a double channel photometer.

The observations were compared with models of the expected S-Z effect 
computed on the basis of X-ray ROSAT and ASCA data of the source.
\hfill\break
From the detected decrement we infer a \comp ~parameter of $(2.60 \pm 0.79 )
~ 10^{-4} $, which is consistent with that computed using the X-ray images.
The 1.2 {\it mm} channel data alone gives rise to a  larger 
\comp ~parameter and this result seems to be unexplained by presence
of known sources in the cluster unless a strong emission from intracluster
dust will be found. Unknown instrument systematics could contribute to this
discrepancy.

We then use ROSAT and ASCA X-ray observations to model
the S-Z surface brightness.
Since the cluster is asymmetrical and probably in a merging process,
modeling is only approximate. The complex morphology of the cluster
has been taken into account by exploring a set of alternative models,
which were used to bracket the associated uncertainty.
We then find
as the global uncertainty on the \comp ~parameter a factor of 1.3.
\hfill\break
We have also considered the effects of the pointing uncertainties of
the ROSAT and SEST Telescopes on the estimated $Y$.

The Combination of the S-Z and the X-ray measurements allows us to
estimate a value for the Hubble constant:
$H_0(q_0=\frac{1}{2})= 53 ^{+ 38} _{- 28} \,{\rm kms}^{-1}{\rm Mpc}^{-1}$,
where the uncertainty is conservative and takes into account
those related to modeling the X-ray plasma halo, the pointing
accuracy and the statistical/calibration/systematics
 uncertainty of the {\rm mm} observations.

As a final exercise we determine the total gas mass from the S-Z measurements
and compare it with that inferred from the X-ray data.
Following Aghanim et al. (1997) we find for $H_0=50 km s^{-1} Mpc ^{-1}$
($q_0=0.5$):

$$M_{gas/SZ} = \frac{(\sqrt{1+z}-1)^2}{(1+z)^3}
\frac{Y}{0.43} \left(\frac{10 keV}{kT_e}\right) h^{-2}
10^{14} M_\odot = 3.15 ~10^{15} M_\odot$$

\noindent
where $Y$ is the integral over the solid angle of the \comp~ parameter
$Y = \int \frac{y d\Omega}{10^{-4} arcmin^2}$.
\hfill\break
The ratio $\frac{M_{gas/Xray}}{M_{gas/SZ}} = 1.11 h_{50}^{-1/2} $
is derived integrating the mass of the X-ray emitting gas using
the hydrostatic, isothermal beta-model with parameters given in table 2
up to the virial radius
($3 h_{50}^{-1} Mpc$). The agreement between the two estimates is excellent,
within the large errorbars, for $H_0=50 km s^{-1} Mpc ^{-1}$.

\vskip 1cm
\section*{Acknowledgements}
The authors are indebted to the ESO/SEST teams at La Silla and in particular
to Peter de Bruin, Peter Sinclair, Nicolas Haddad and Cathy Horellou.
This work has been partially supported by the P.N.R.A. (Programma
Nazionale di Ricerche in Antartide). P.A. warmly thanks ESO for hospitality
during 1995, when part of this work was carried out.
We thank Yasushi Ikebe for his help in the reduction  
of the ASCA data. 
We are grateful to the editor, E.L. Wright, and to an unknown referee,
whose suggestions and comments helped in improving the paper.
We made use of the {\it Skyview} Database,
developed under NASA ADP Grant NAS5-32068.

\newpage

\appendix

\section{A brief description of the photometer}

The photometer is a 
two-channel device covering the frequency bands 129-167 GHz and 217-284 
GHz. It uses two Si-bolometers cooled to 0.3 K by means of
a single stage $^3$He refrigerator.
Radiation coming from the telescope is focussed by a PTFE lens
and enters the cryostat through a
polyethylene vacuum window and two Yoshinaga edge filters one
cooled at 77 K and the other cooled at 4.2 K. A dichroic mirror
(a low-pass edge filter: beam-splitter) at 4.2 K
divides the incoming radiation between two f/4.3 Winston Cones cooled
at 0.3 K located orthogonal each other. 

\subsection{Optics response}

The wavelength ranges are defined by two interference filters centred at
1.2 and 2 {\it mm} cooled at 4.2K with bandwidths 350 and 560 $\mu m$
respectively.
Their rejection factor has been estimated to be better than 10$^{-6}$.
At the Winston cone entrance a further Yoshinaga type filter cooled at 0.3 K
is installed.
Figure A1 shows the measured transmission spectra of the two trains of filters:
form the vacuum window to the final Yoshinaga at 0.3 K.

\newpage
\bigskip
\begin{center}
{{\bf Table 1.} Performances of the photometer at focus}
\end{center}
\begin{center}
\begin{tabular}{cccccccccccc}
\hline
\hline
&& &$\lambda _c$ & $\Delta \lambda$ & FWHM
 & \multicolumn{1}{c} {noise}
 & \multicolumn{1}{c} {Responsivities}
 & \multicolumn{1}{c} {NET$_{ant}$}
 & \multicolumn{1}{c} {$({\Delta T\over T})_{therm}$} & &\\
&& & ($\mu m$) & ($\mu m$) & ($^\prime$) &
\multicolumn{1}{c}{(nV/$\sqrt{Hz}$)}
 & \multicolumn{1}{c}{($\mu V/K$)}
& \multicolumn{1}{c}{(mK/$\sqrt{s}$)} & \multicolumn{1}{c}{(1s)} & &\\
\hline
\hline
&&&1200 & 360 & 44 & 45 & 3.0 & 10.6 & 0.0142 &&\\
&&&2000 & 580 & 46 & 31 & 1.4 & 15.6 & 0.0099 &&\\
\hline
\hline
\end{tabular}
\end{center}

\bigskip
\bigskip
\begin{center}
{{\bf Table 2.} Results of the $\beta$-model fits to the surface
brightness profiles of the ROSAT PSPC and HRI observations}
\end{center}
\begin{center}
\begin{tabular}{ccccccccc}
\hline
\hline
&& &Data  & $S_0$  & {\rm core~radius} & {$\beta$} & $r_c$ &\\
&& & & ({\it cts s$^{-1}$ arcmin$^{-2}$}) & {\it arcmin}  & & ($h_{50}^{-1}$
Mpc) &\\
\hline
\hline
&&& {\rm Eastern ~component}  & 0.027 & 1.23 & 0.7  & 0.406 &\\
&&& {\rm Western ~component}  & 0.046 & 0.26 & 0.49 & 0.086 &\\
\hline
\hline
\end{tabular}
\end{center}

\newpage
\begin{figure}[ht]
\plotone{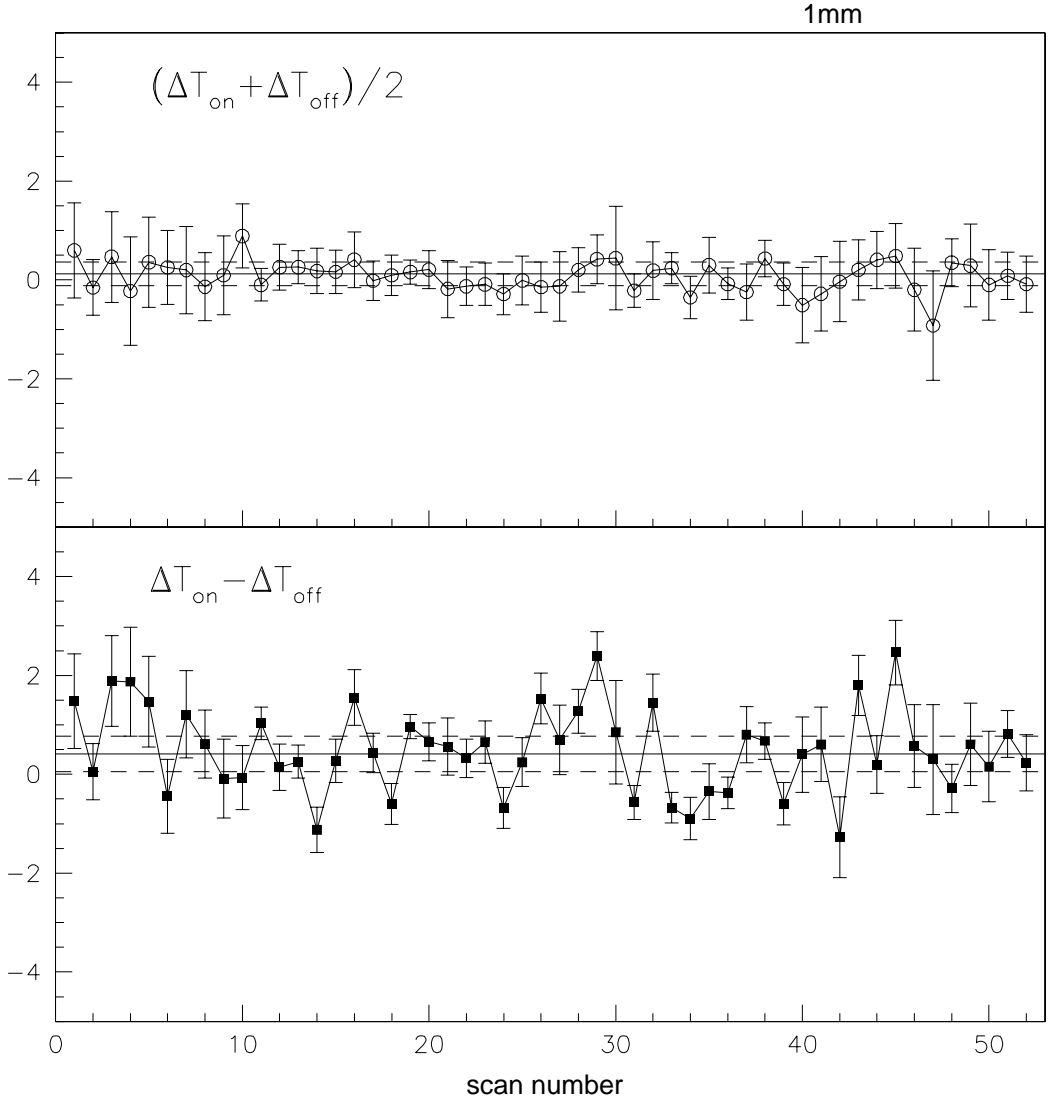}
\caption[]
{(a) Differential antenna temperatures at 1 {\it mm} for RXJ0658-5557. 
The cluster signals are estimated by subtracting 
$ \Delta T_{SZ} = \Delta T_{ON} - \Delta T_{OFF} $ and are reported in the 
lower panel. The upper panel shows the sum 
$ \frac{\Delta T_{ON} + \Delta T_{OFF}}{2} $ which is sensitive to any possible 
systematics. 
The maximum likelihood values of $ \Delta T_{SZ} $ is shown as a solid line 
while the dashed lines correspond to $\pm 3 \sigma$ confidence interval.}
\label{onoff1}
\end{figure}
\begin{figure}[ht]
\plotone{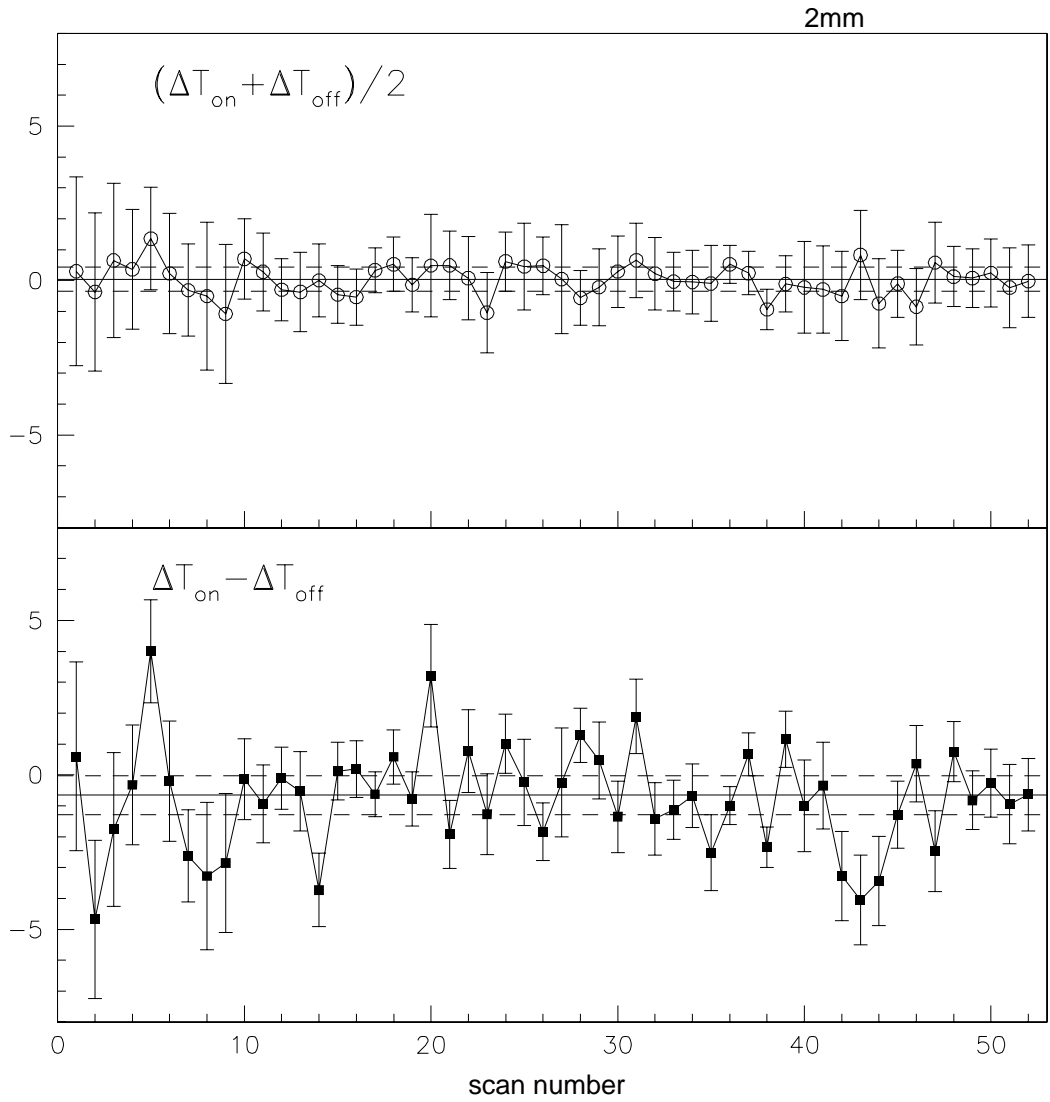}
\caption[]
{(b)
The same as Fig. 1a but for the 2 {\it mm} channel.}
\label{onoff2}
\end{figure}
\begin{figure}[ht]
\plotone{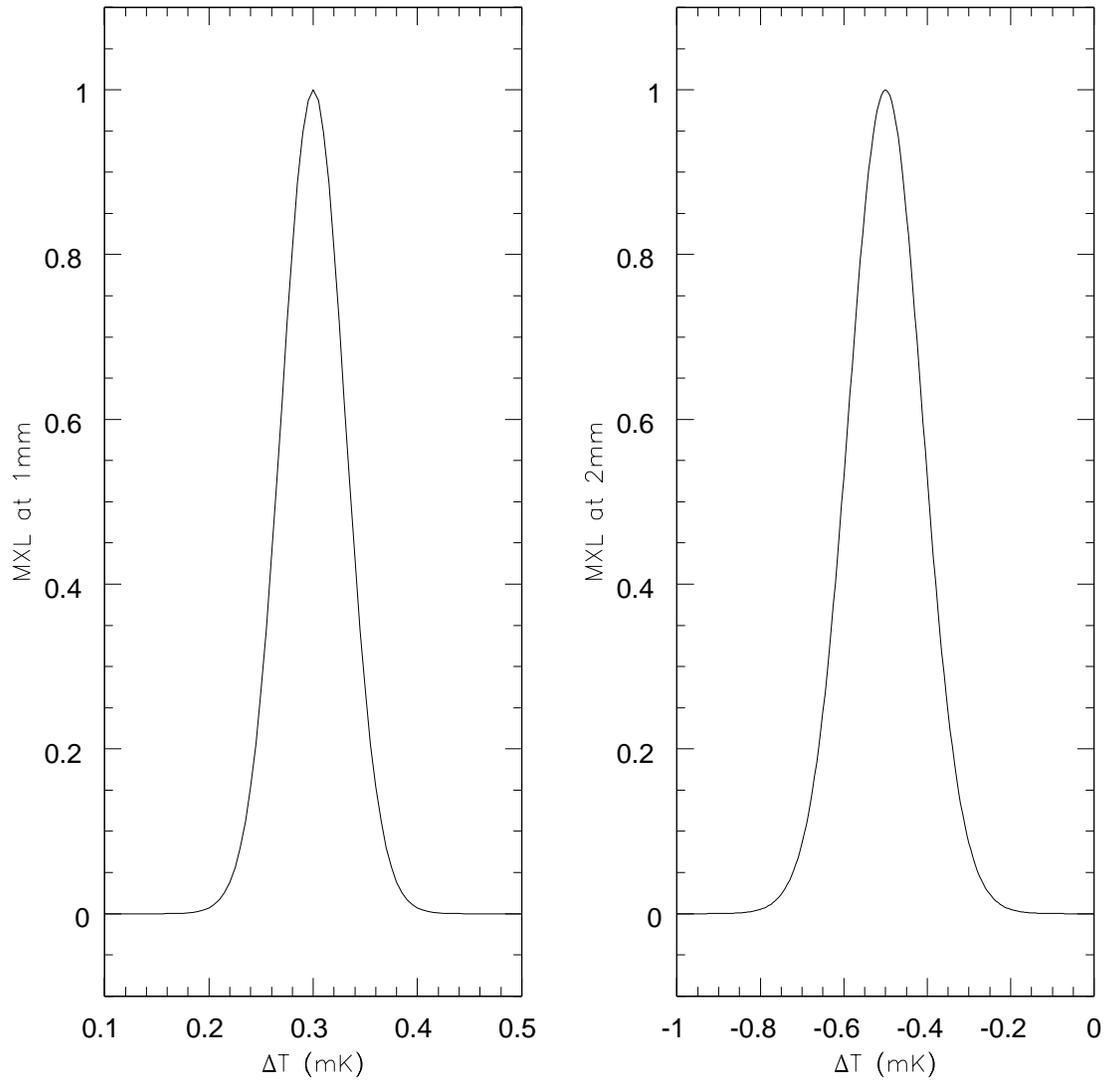}
\caption[]
{Maximum likelihood curves of the data reported in Figure 1. Abscissa values
are in antenna temperatures.}
\label{mx}
\end{figure}
\begin{figure}[ht]
\plotone{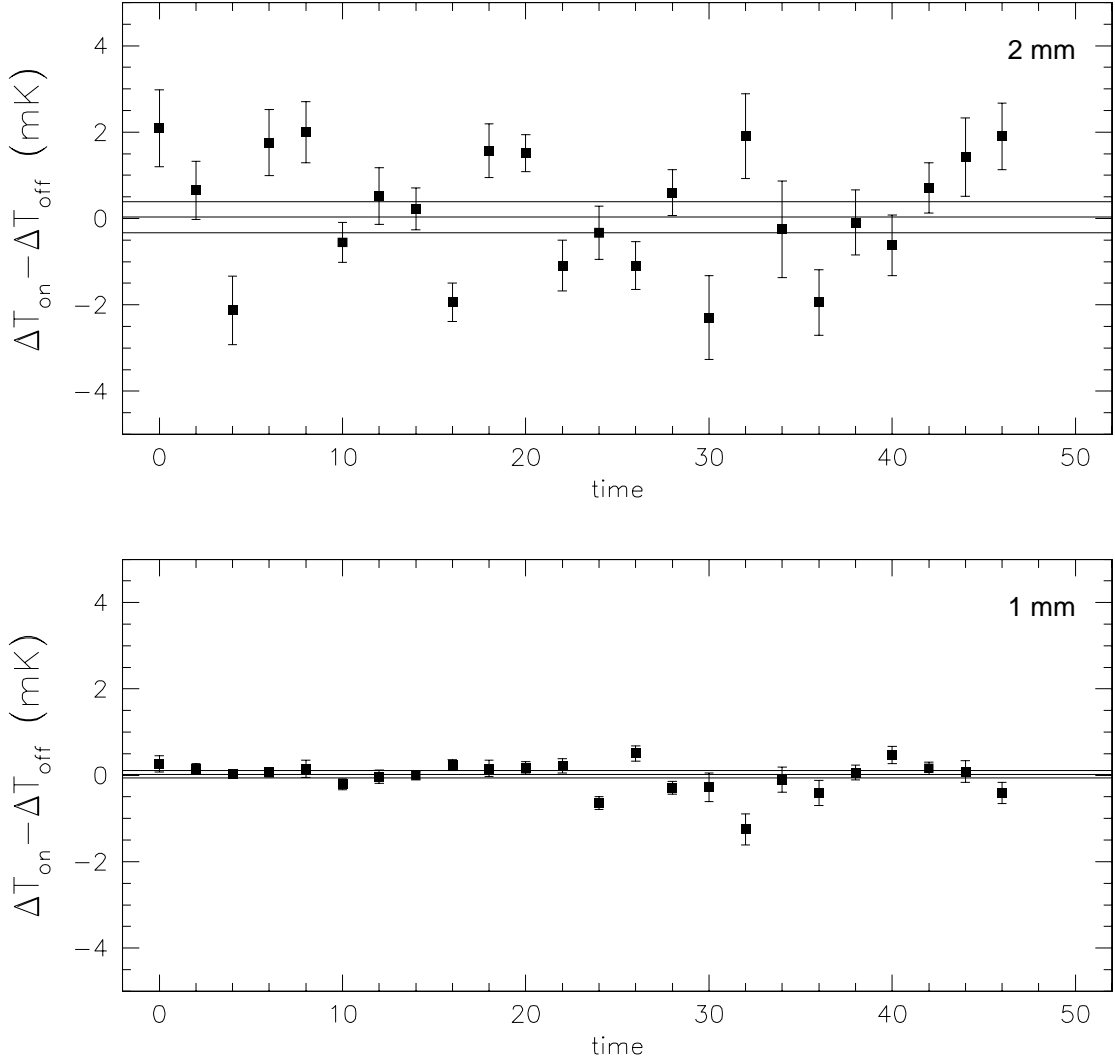}
\caption[]
{Antenna temperature differences,
$\Delta T _{ON} - \Delta T _{OFF}$,
 taken with the photometer window covered with a metal sheet.
The same observing procedure, ON and OFF the source,
was applied with this configuration to test the photometer systematics.
It is clear that there are no spurious signals generated from the instrument.
Solid lines correspond to the averages computed over the plotted data
$\pm$ 3 $\sigma$. Note that the total integration time in this case is
three times shorter than that in figure 1.}
\label{ms}
\end{figure}
\begin{figure}[ht]
\plotone{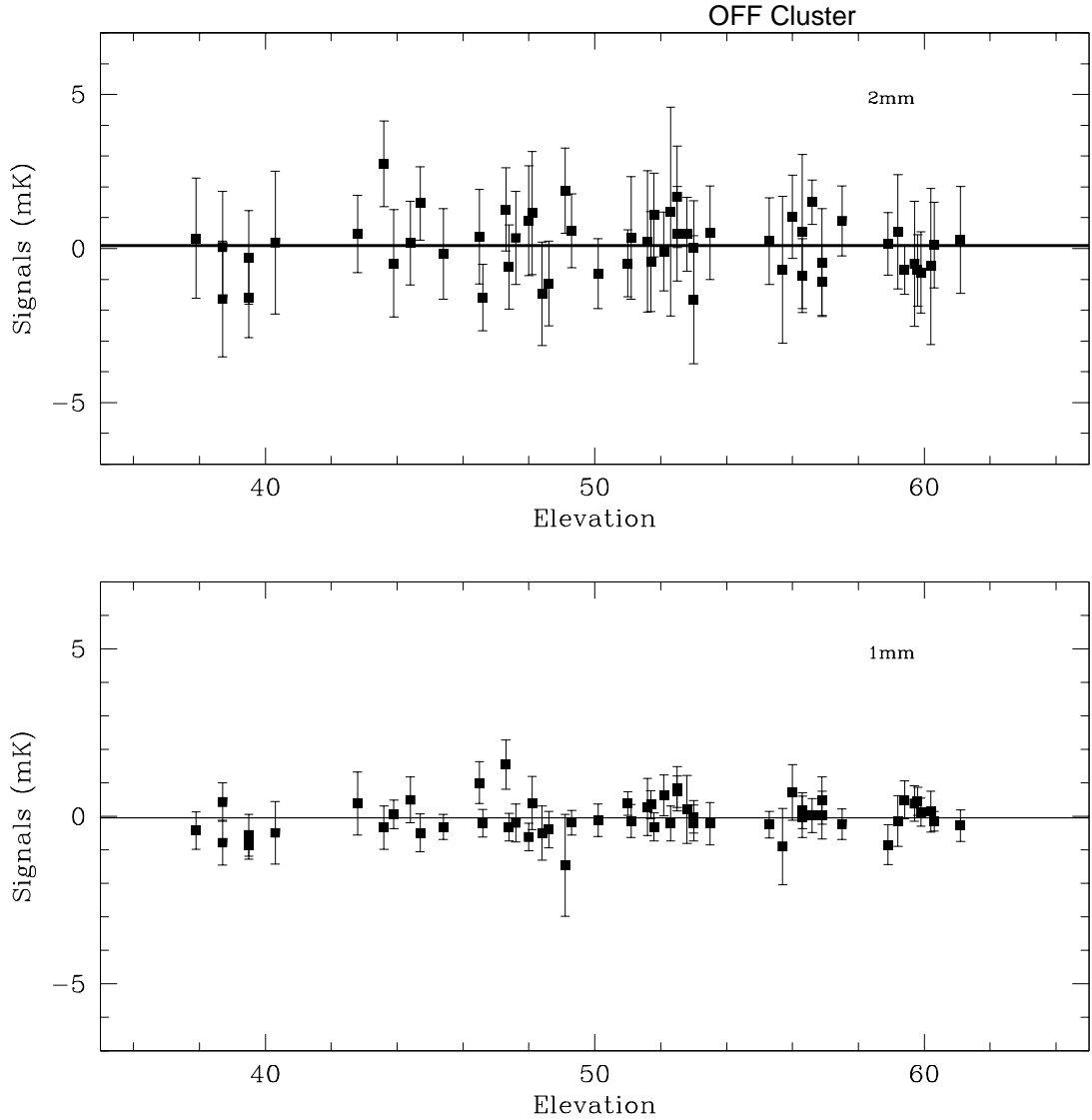}
\caption[]
{Antenna Temperature values taken OFF the source (on blank sky at position
15m ahead in R.A.) as a function of the elevation of the source.
Solid lines correspond to the averages in antenna temperatures:
$0.09\pm0.06$ mK (2 mm), $-0.03\pm0.02$ mK (1mm)}
\label{bs}
\end{figure}
\begin{figure}[ht]
\plotone{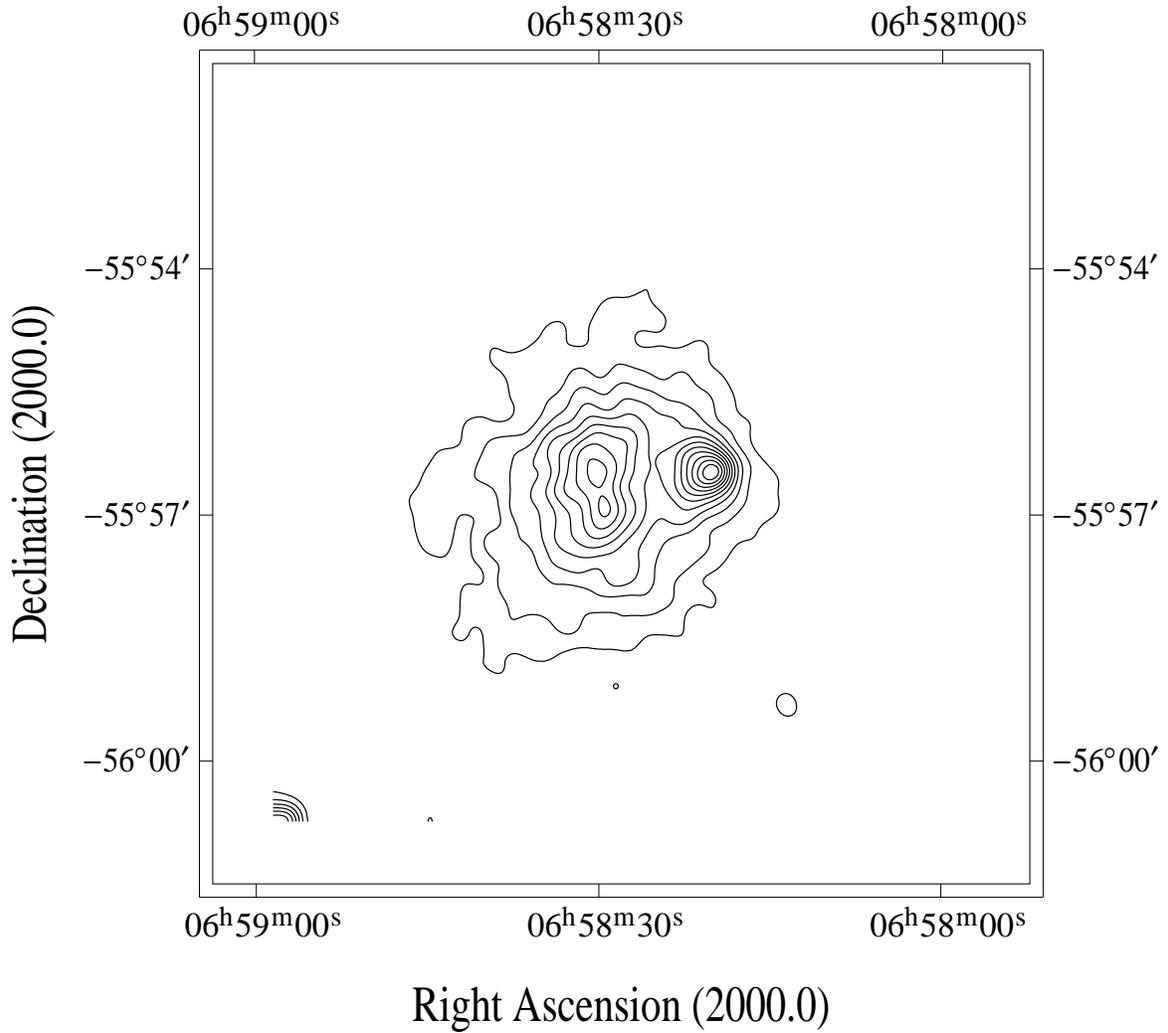}
\caption[]
{X-ray image of the RXJ0658-5557 cluster observed
with the ROSAT/PSPC.
The contour levels are logarithmically spaced with the peak
brightness corresponding to $9.4 \times 10^{-2}\,{\rm cts\,arcmin}^{-2}$.}
\label{X}
\end{figure}
\begin{figure}[ht]
\plotone{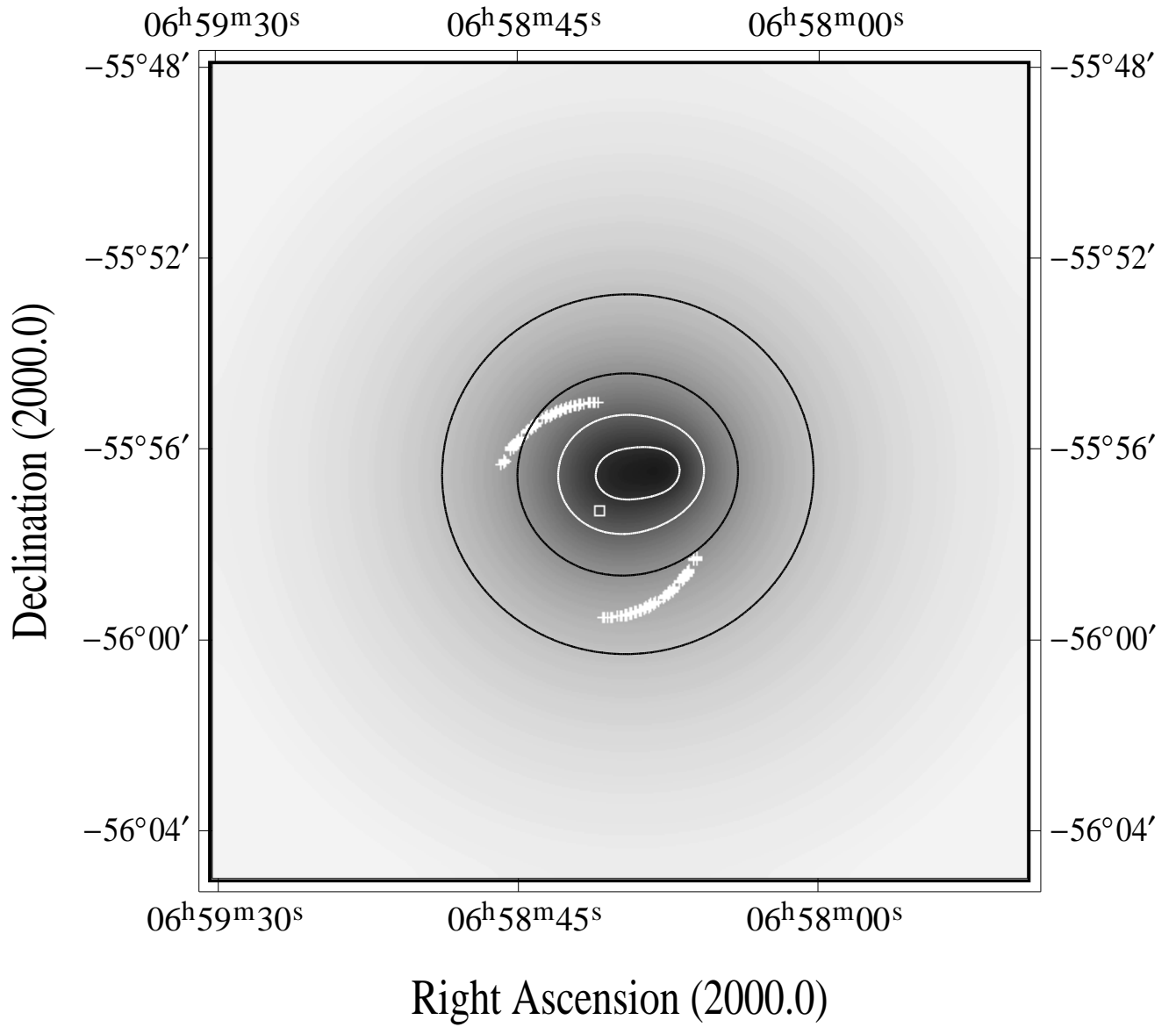}
\caption[]
{The convolved S-Z surface brightness predicted with the
X-ray model described in \S 4.1. Superimposed are the northern
and southern arcs of the reference beams (crosses) and the
position of the main beam (open square).}
\label{SZ}
\end{figure}
\begin{figure}[ht]
\plotone{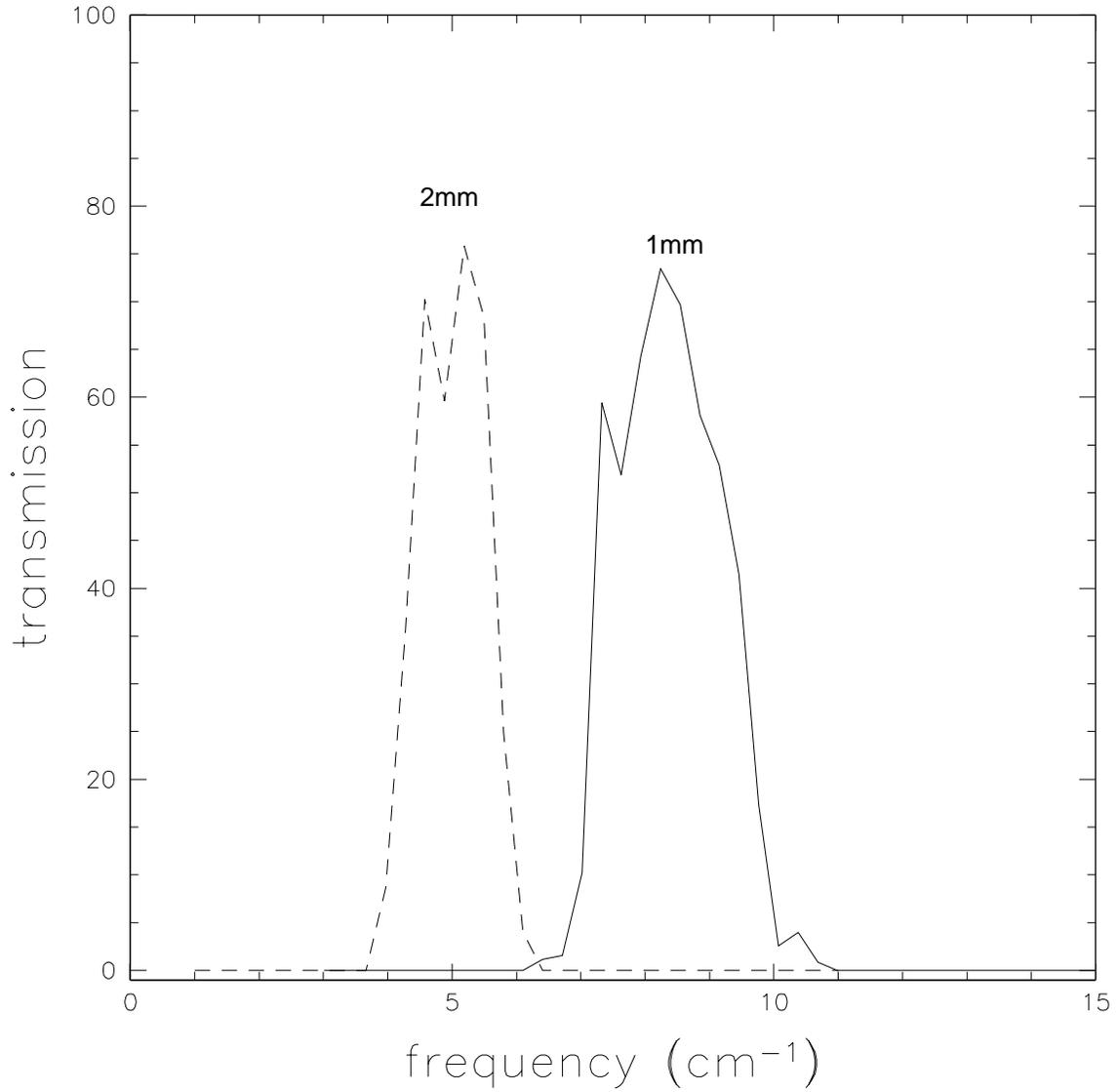}
\caption[]
{Transmission spectra of the two optical trains defining the wavelength
ranges of the two channels at 1 {\rm mm} ($\sim$ 8.5 $cm^{-1}$) and
at 2 {\rm mm} ($\sim$ 5 $cm^{-1}$).}
\label{appen}
\end{figure}

\end{document}